\documentclass[12pt]{article}

 \usepackage{caption}
 \usepackage{subcaption}
 \usepackage{float}

 \usepackage{appendix}
 \usepackage{hyperref}

\usepackage{amssymb}
\usepackage{euscript}

\usepackage[utf8]{inputenc}
\usepackage[T1]{fontenc}
\usepackage{amsmath}
\usepackage{amsfonts}
\usepackage{amssymb}
\usepackage{latexsym}
\usepackage[english]{babel}
\usepackage[dvips]{graphicx,color}
\usepackage{latexsym}
\usepackage{graphicx}
\usepackage{graphics}
\usepackage{pstricks}
\usepackage{authblk}

\newcommand\be{\begin{equation}}
\newcommand\ee{\end{equation}}
\newcommand\ben{\begin{equation*}}
\newcommand\een{\end{equation*}}
\newcommand\ba{\begin{eqnarray}}    %eqnarray
\newcommand\ea{\end{eqnarray}}      %eqnarray

\title{Which quantum states are dual to classical spacetimes?}
\author[$\dagger$]{Marcelo Botta-Cantcheff}
\author[$\dagger$]{Pedro J. Martínez}
\affil[$\dagger$]{{\it  Instituto de F\'\i sica de La Plata, CCT La Plata - CONICET \& 

Departamento de F\'\i sica - Universidad Nacional de La Plata,  C.C. 67, 1900 La Plata, Argentina}
~

{\tt E-mail:} botta@fisica.unlp.edu.ar, martinezp@fisica.unlp.edu.ar}
\date{}

\begin{document}

\begin{titlepage}

\maketitle

\begin{abstract}

It is commonly accepted that states in a conformal field theory correspond to classical spacetimes with Anti-de-Sitter asymptotics.
In this work we give a prescription for the CFT states with a dual classical spacetime and, using basic holographic rules,
show that they are holographically connected to coherent states in the large-N limit, or by considering linearized perturbations.
 
 We also point out implications in the spacetime emergence mechanism, for instance,
%the eigenstates of the Hamiltonian have not a good dual spacetime, and 
the (entangled) state dual to the black hole should be properly described as a quantum superposition of products of these states. This also complements the quantum interpretation of the Hawking-Page transition.

\end{abstract}

\thispagestyle{empty}
%\tableofcontents
\end{titlepage}

\textbf{Introduction}
\vspace{.4cm}

The AdS/CFT correspondence represents the paradigmatic case of gravity/gauge duality where spacetime with fixed (AdS) asymptotics can be defined
 as emergent from a ordinary quantum field theory defined on its conformal boundary \cite{adscft}. However, we do not understand the mechanism of this emergence in depth.

It is widely accepted that states in the CFT Hilbert space are dual to classical asymptotically AdS (aAdS) spacetimes but we do not know yet which aspects of the classical geometries are encoded in the states or how to read off such aspects. Moreover, it is unclear which CFT states actually are dual to some type of classical geometry. An illuminating observation was made by Van Raamsdonk some years ago \cite{vanram}, who argued that classically connected spacetimes correspond to \emph{entangled} states in the CFT, and furthermore, that they are a quantum superposition of basis states supposedly dual themselves to some kind of aAdS spacetimes. This set up has been used in further developments \cite{collapse, VRrindler}.

The present work is devoted to study the (disentangled) microstates that compose the spacetime geometry as a quantum superposition, and whether they are dual to geometries with usual classical properties.

The more fundamental goal is to construct a holographic map that work in the same way that the paradigmatic cases, namely, the CFT vacuum  $ |0 \rangle \mapsto $ AdS, and the thermal TFD state $|0 (\beta)\rangle \mapsto$ AdS-Black hole. In fact, based on well established prescriptions \cite{SvRC,SvRL,BDHM,BDHM2},  and the Skenderis $\&$ van Rees suggestion to construct excitations, the results of \cite{us} gave us a clue about how to generalize this mechanism to other (excited) states in the same spirit.

In line with this, and previous literature 
\cite{cosascoherentes}, we will argue that quantum coherence is also an essential ingredient for the emergence of classical spacetimes, and as a result, that the (eigen-energy) microstates $\left|E_n\right\rangle_1 \otimes \left|E_n\right\rangle_2$ that superpose to form an AdS Black Hole do not correspond to any (disconnected) pair of aAdS geometries.

This letter is organized as follows: we first review the original proposal \cite{vanram}, understanding a BH as a quantum superposition of states living in two CFT theories. We then review the construction \cite{us}, extending the original SvR prescription \cite{SvRC,SvRL} to build the CFT states with a well defined dual geometry associated, and then argue that, in the large $N$ limit the eigen-energy basis cannot be associated to any classical aAdS geometry, although the holographic states form an overcomplete basis and any state can be described as a quantum superposition of them. In the last part, we study the  AdS black hole with matter fields and show that it fits into our prescription. The final observations and remarks are collected at conclusions.

\vspace{.4cm}
\textbf{The emergent AdS Black Hole}
\vspace{.4cm}

The standard interpretation is that the exact bulk geometry AdS$_{d+1}$ corresponds to the fundamental state $\left|0\right\rangle$ of the CFT Hilbert space ${\cal H}$ defined on its conformal boundary
$S^d \times {\mathbb R}$, and general classical aAdS spacetimes should be dual to certain excited states.

Let us consider now two (non-interacting) identical copies 
%CFT$_{i}$, $i=1,2$
of this CFT (labeled by a subindex $1,2$). 
%\textbf{cambio}
The asymptotically $AdS_{d+1}$ spacetime with a eternal black hole corresponds to the (entangled) state \cite{eternal}:
\be \left|0(\beta)\right\rangle = \sum_n \, \frac{e^{-\frac{\beta}{2} \, E_n}}{Z^{1/2}}\left|E_n\right\rangle_1 \otimes \left|E_n\right\rangle_2
~\in{\cal H}_1\otimes{\cal H}_2~~,~~\beta\equiv(k_B T)^{-1}\label{BHstate}, \ee
% where the $|...\rangle\!\rangle$ denotes a ket that belongs to the product of the CFT Hilbert spaces ${\cal H}_1\otimes{\cal H}_2$, 
where the $\left|E_n\right\rangle$ are a complete basis of eigenstates of the CFT Hamiltonian $H$, and $E_n$ are its eigenvalues.
This describes a thermal state of the CFT system at temperature $T$ in the \emph{thermofield dynamics} (TFD) formalism \cite{tu,ume1,ume2}.
Recall also that the state \eqref{BHstate} is invariant under the action of the combination $ H_1 - H_2$. 

\begin{figure}[t]\centering
\includegraphics[width=.9\linewidth] {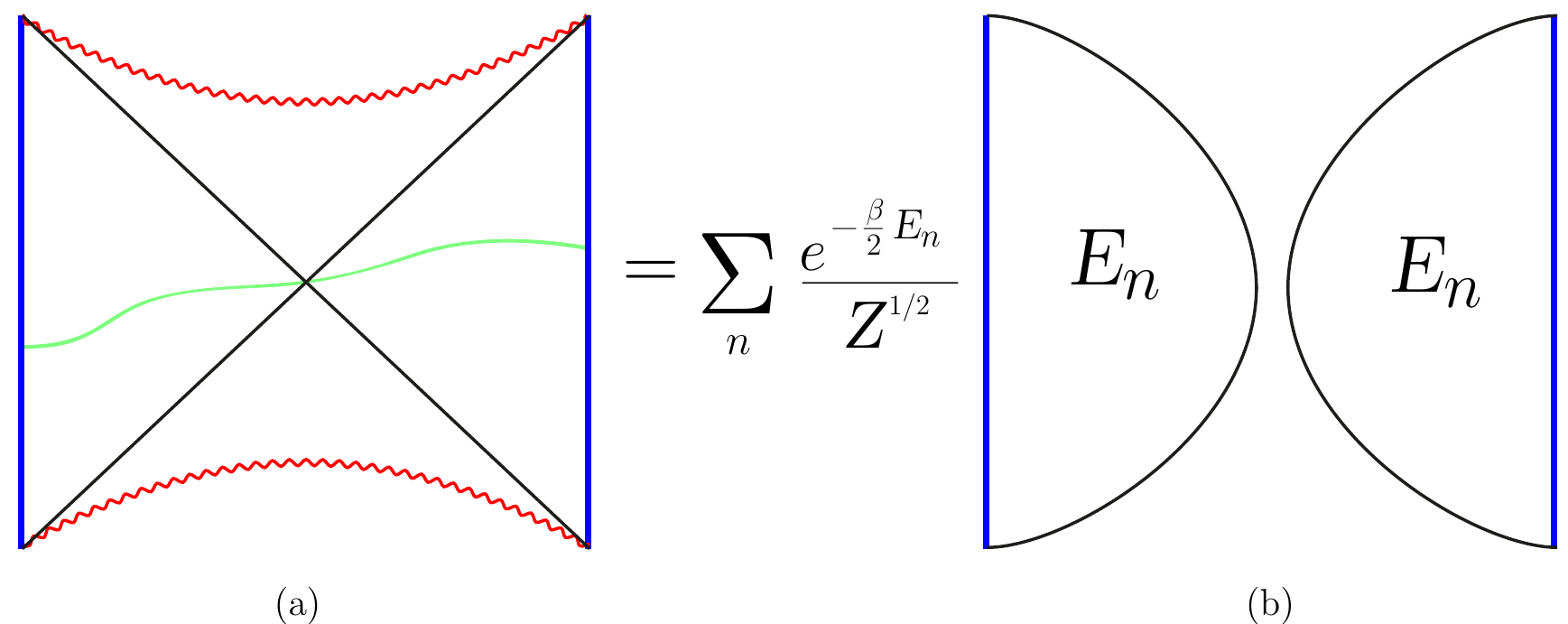}
\caption{\small{ (a) Penrose diagram of a maximally extended AdS-black hole. The green line is a connected spacial slice representing the entaglement between the CFT's. (b) We schematically show the interpretation of \cite{vanram}, where the resulting state (\ref{BHstate}) is a linear combination of states $|E_n\rangle_1 \otimes |E_n\rangle_2$ supposedly dual to aAdS spacetimes. The blue lines represent the non-interacting CFT theories on the two asymptotic boundaries.}}
\label{vanramfig}
\end{figure}

One of the arguments of \cite{vanram} was precisely based in the structure of this state, which describes a classically connected spacetime (fig. \ref{vanramfig}a) with
two aAdS regions causally separated by an event horizon \cite{galloway} using a quantum superposition of
states $\left|E_n\right\rangle_1 \otimes \left|E_n\right\rangle_2$ (fig. \ref{vanramfig}b), which,
%\emph{if} correspond to classical geometries, should be disconnected aAdS spacetimes.
 \emph{if} have a classical geometric dual, %\textbf{cambio} %they 
 should correspond to a pair of disconnected aAdS spacetimes.

The argument for it is that disentangled states
$|\Psi\rangle_1 \otimes |\Upsilon \rangle_2 \in {\cal H}_1 \otimes {\cal H}_2 $ of two systems CFT$_1$ and CFT$_2$
that do not interact in any way, must describe
two completely separate physical systems. Then if $\left|\Psi\right\rangle_1$ is dual to one aAdS
spacetime and $\left|\Upsilon\right\rangle_2$ is dual to some other spacetime, the product state must be a geometry dual to the
disconnected pair of spacetimes (see Fig \ref{vanramfig}b).
In particular, the ground state is $\left|0\right\rangle_1 \otimes \left|0\right\rangle_2$, that correspond to two disconnected globally AdS spacetimes \cite{vanram}.

In order to turn this interpretation useful for holographic quantum gravity one should better understand which is the geometric content of these (micro)states, and whether they are holographic themselves.
Below, we will propose a suitable definition of the states in each CFT copy that are holographically dual to aAdS spacetimes and show that the elements $\left|E_n\right\rangle_1 \otimes \left|E_n\right\rangle_2$ do not fit into this definition. The immediate implication is that one should find another way of representing the $\left|E_n\right\rangle$'s in the gravity side, perhaps, in terms of other type of non-conventional \textit{microscopic} geometry rather than a smooth $d$-dimensional spacetime, or more conventionally, as a quantum superposition of good holographic states. The present approach follow this line.

\vspace{.4cm}
\textbf{CFT states with spacetime dual}
\vspace{.4cm}

\begin{figure}[t]\centering
\begin{subfigure}{0.49\textwidth}\centering
\includegraphics[width=.9\linewidth] {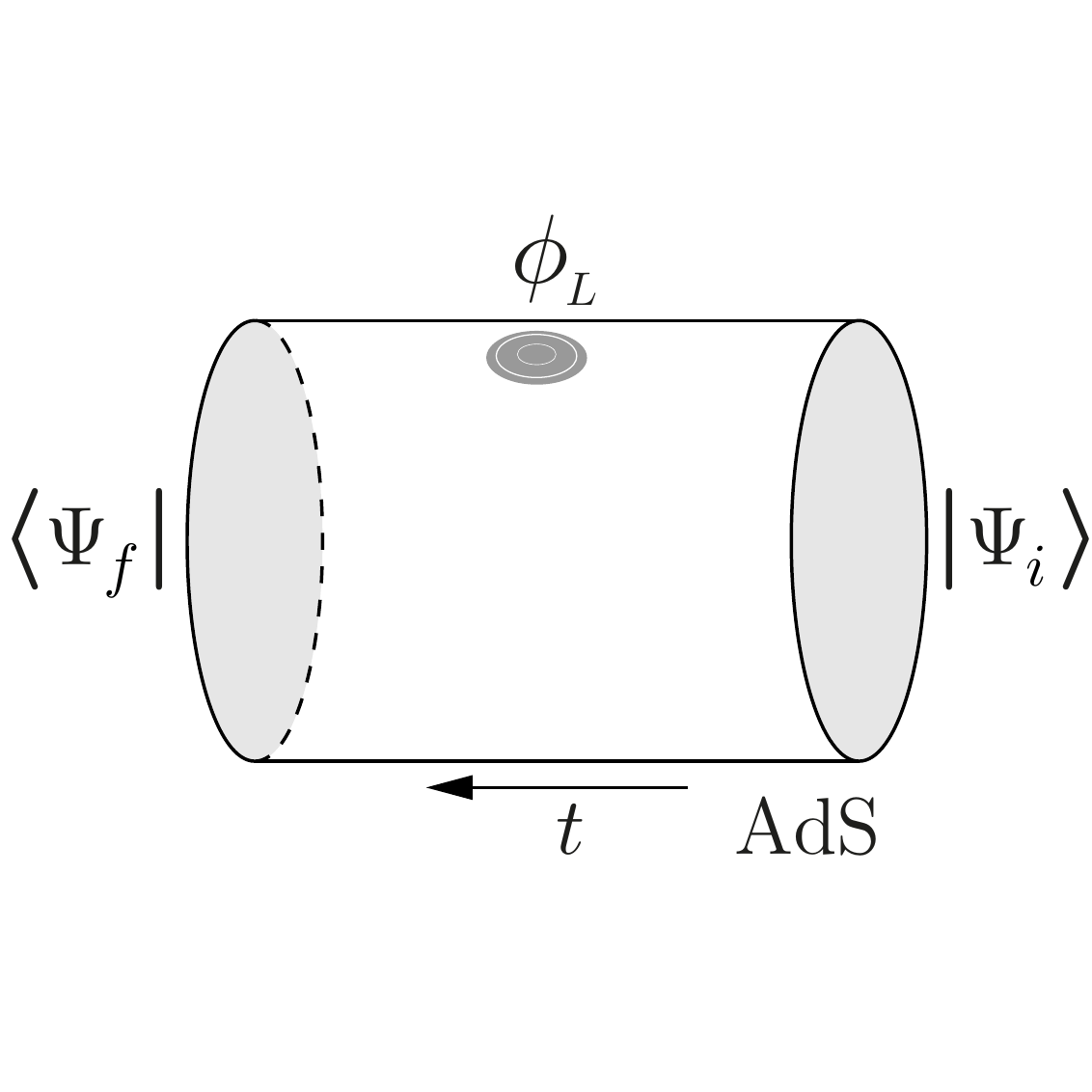}
\caption{}
\end{subfigure}
\begin{subfigure}{0.49\textwidth}\centering
\includegraphics[width=.9\linewidth] {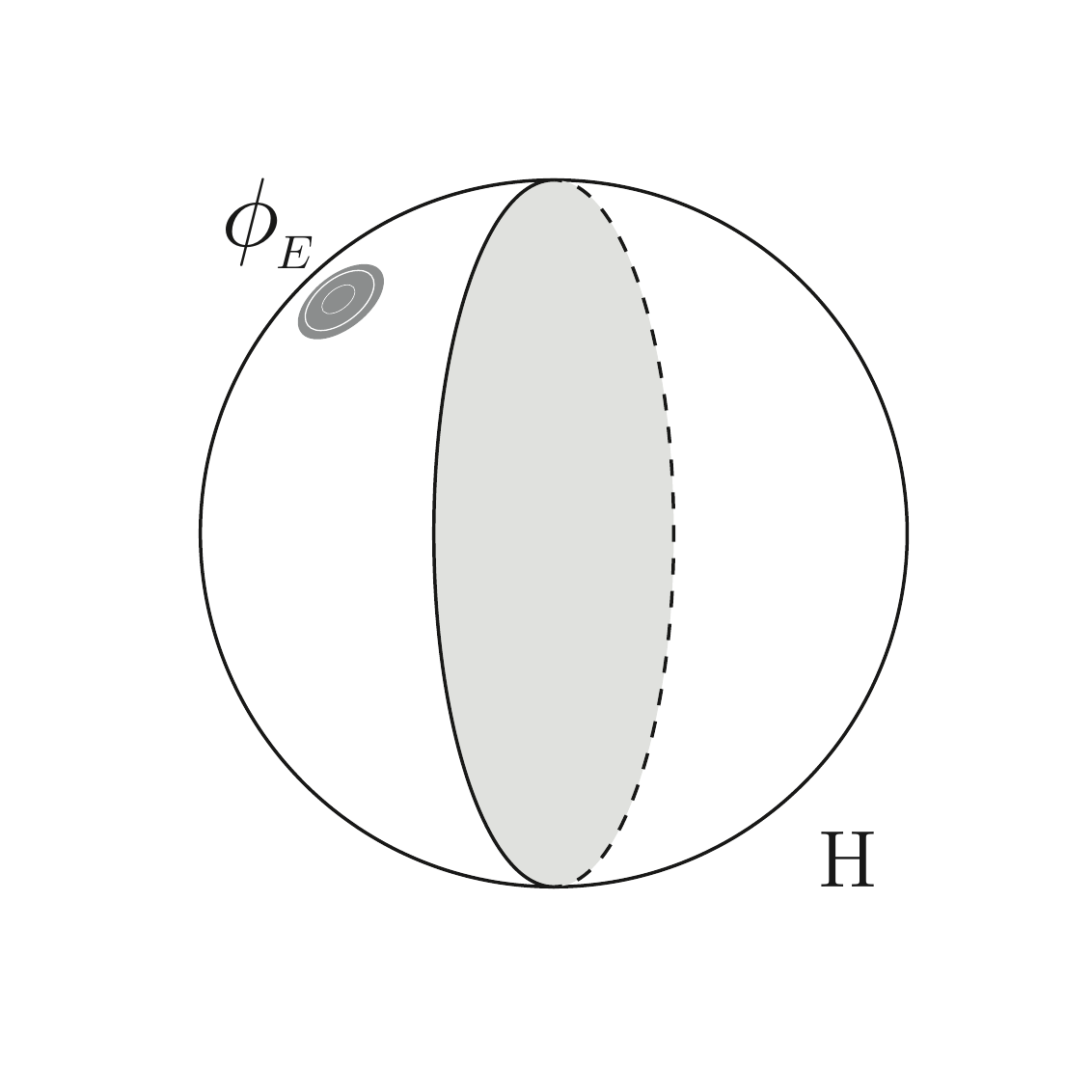}
\caption{}
\end{subfigure}
\caption{\small{(a) Lorentzian AdS with boundary condition $\phi_L$. We also schematically depict the initial and final states. (b) Euclidean AdS (hyperbolic space).}}
\label{Sources}
\end{figure}

\begin{figure}[t]\centering
\begin{subfigure}{0.49\textwidth}\centering
\includegraphics[width=.9\linewidth] {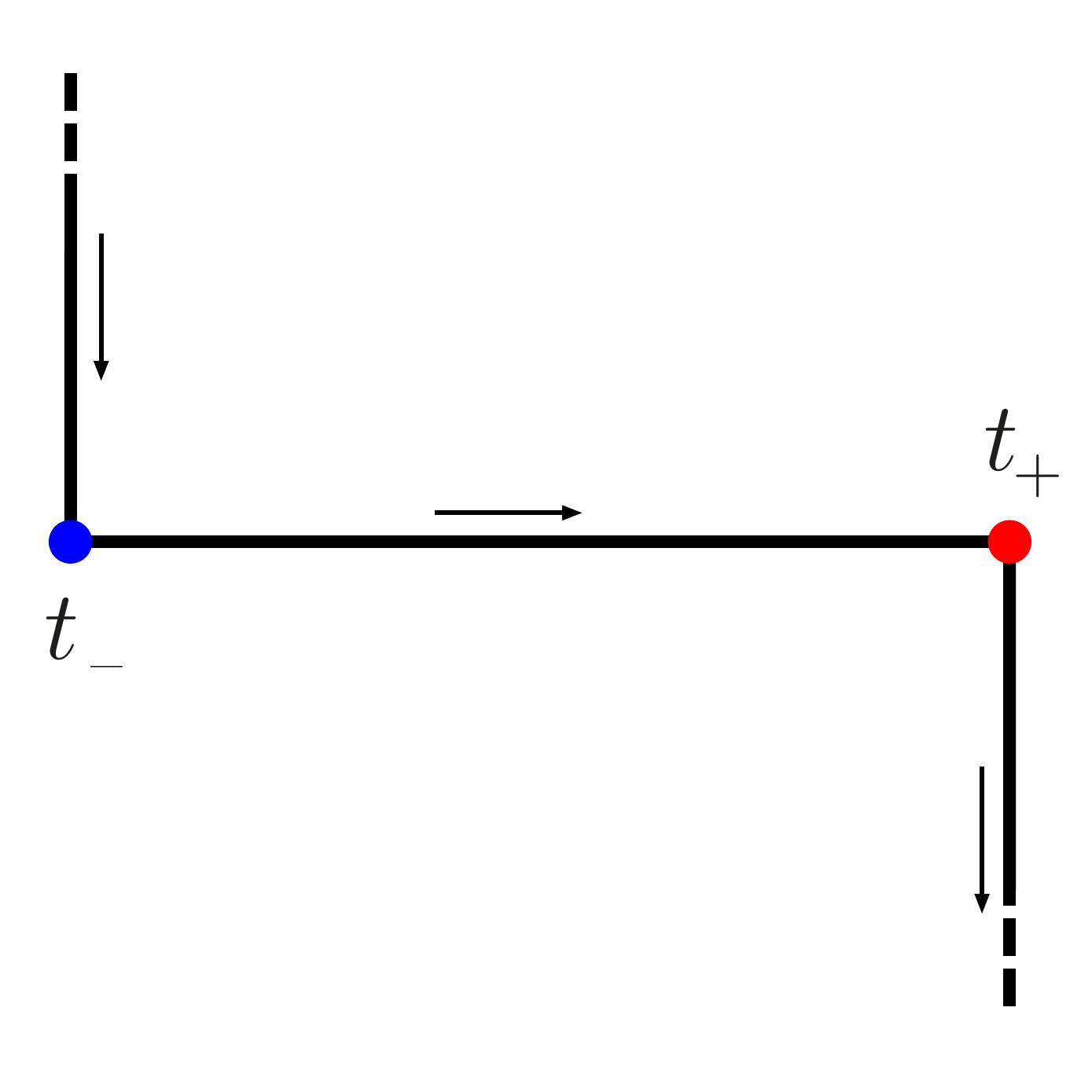}
\caption{}\label{SvRa}
\end{subfigure}
\begin{subfigure}{0.49\textwidth}\centering
\includegraphics[width=.9\linewidth] {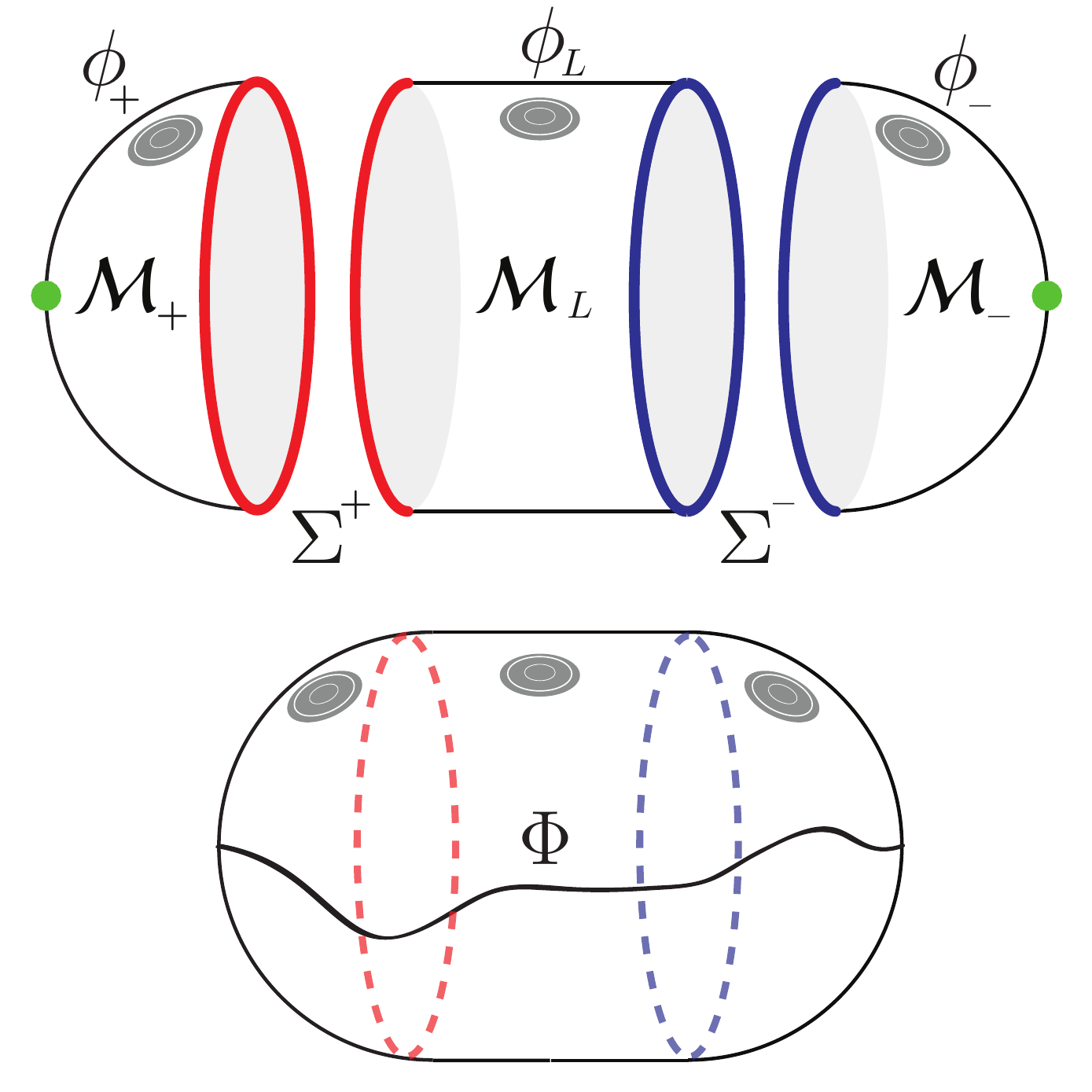}
\caption{}\label{SvRb}
\end{subfigure}
\caption{\small{(a) CFT contour in complex t-plane showing real time interval $( t_- , t_+)$. (b) Geometric dual to the contour obtained by gluing together Lorentzian ${\cal M}_L$ and Euclidean  sections (two halves) of the AdS spacetime ${\cal M_\pm}$. We depict the gluing surfaces $\Sigma^\pm$ and the green points correspond to $\tau=\pm\infty$. On the bottom we show the glued geometry with a generic classical configuration.}}
\label{SvR}
\end{figure}

The prescription \cite{SvRC,SvRL} allows the calculation of time ordered $n$-point
correlation functions of local CFT operatos $\mathcal{O}$ in AdS/CFT (fig. \ref{Sources}a). It can be
understood as the real time version of the standard prescription \cite{GKP,W} (fig. \ref{Sources}b), and corresponds to the geometry shown in figure \ref{SvRa}.

\begin{equation}
\label{SvR:Main:SvR-Def}
\langle 0 | \, e^{-i \int_{\partial_r \mathcal{M}_L} \mathcal{O} \phi_L }\,  | 0 \rangle
= e^{i S^0_L[\phi_L;\phi_{\Sigma^-},\phi_{\Sigma^+}] - S^0_-[0;\phi_{\Sigma^-}] - S^0_+[0;\phi_{\Sigma^+}]}\,.
\end{equation}

The lhs gives the generating function of time ordered correlation functions of a scalar operator $\cal O$ in a Lorentzian CFT that lives in the
timelike conformal boundary $\partial_r \mathcal{M}_L= S^d \times {\mathbb R}$ of the spacetime $\mathcal{M}_L$. In the rhs, $S^0_L[\phi_L;\phi_{\Sigma^-},\phi_{\Sigma^+}]$
is the Lorentzian on-shell action for a bulk field $\Phi_L$ which takes boundary values $\phi_{\Sigma^\pm}$ on the spacelike boundaries
$\Sigma^{\pm}\equiv\partial_t{\cal M}_L$ and  $\phi_L$ over $\partial_r \mathcal{M}_{L}$. The exponents $S^0_{\pm}[0;\phi_{\Sigma^{\pm}}]$
are the bulk field on shell actions on the Euclidean sections $\mathcal{M}_{\pm}$ for boundary values $\phi_\pm=0$ on $\partial_r \mathcal{M}_{\pm}$
and $\phi_{\Sigma^\pm}$ on $\Sigma^{\pm}$ .
It is worth noticing that (\ref{SvR:Main:SvR-Def}) implicitly assumes the bulk fields, and its conjugated momenta,
to be continuous through the $\Sigma^\pm$ gluing surfaces.

This recipe follows from a complete quantum treatment where the r.h.s. is a path integral, and where one can also consider \emph{excited} CFT initial/final states by giving smooth non-vanishing boundary conditions $\phi_\pm $ over $\partial_r \mathcal{M}_{\pm}$ for all finite value of the euclidean time\footnote{They vanish at $\tau = \pm \infty$, represented as green dots in fig. \ref{SvRb}.} (see refs \cite{us} and \cite{SvRC} for details). In this context, in ref. \cite{us}, it was explicitly shown that the (initial) excited CFT states are precisely given by\footnote{Where $| 0 \rangle $ denotes the ground state, whose holographic dual is the globally AdS spacetime. These can also be constructed on other vacua as the thermal one, associated to eternal AdS black holes, but this will be studied later on.}
\be\label{estadoinicial-resultado}
 | \Psi^{\phi_-} \rangle = \, e^{-\int_{\partial_r {\cal M}_-}  \, \mathcal{O}\, \phi_-} \,  |0 \rangle\,.
\ee
These states, projected on a complete bulk configuration basis $\phi_{\Sigma^\pm}$ on spacelike surfaces $\Sigma^{\pm}$  correspond
to wave functionals that can be expressed as euclidean path integrals in the gravity side,
\begin{equation}
\label{wavef-g}
\Psi^{\phi_{-}}[\phi_{\Sigma^-}] \equiv \int [\mathcal{D}\Phi]_{(\phi_{\Sigma^-}\,,\,\phi_{-})} \,  e^{-S^{}_-[\Phi]}\,\, ,
\end{equation}
with boundary conditions $\phi_{\Sigma^-}$ and $\phi_{-}$ on $\Sigma^-$ and $\partial_r {\cal M}_-$ respectively. Moreover, in the large-$N$ limit one can use the saddle point approximation and they can be associated to a single classical solution for the bulk fields as
\begin{equation}
\label{wavef-g-largeN}
\Psi^{\phi_{-}}[\phi_{\Sigma^-}] \sim \,  e^{-S^{0}_-[\phi_{\Sigma^-},\phi_{-}]}\,\,.
\end{equation}
 This generalizes the Hartle-Hawking construction of wave functionals \cite{HH} to excited states in AdS quantum gravity \cite{us, SvRC}.
 Obviously this analysis can be properly extended to include the gravitational degrees of freedom, although in our arguments here we consider only a scalar field that probes the spacetime geometry. In fact, the existence of classical fields implies many classical properties of the background manifold $\mathcal{M}$, such as connectivity and differentiability.

Summarizing, we conclude that 

states \eqref{estadoinicial-resultado} can be holographically associated to well defined classical
geometric configurations. For a state of the quantum (bulk) theory \eqref{wavef-g}, there is a set of histories (geometries)
associated. Then in the large-N limit, there is a unique
 path describing a classical (euclidean) spacetime with matter fields %\textbf{corregir:} 
 $(g^{cl}_{\mu \nu} , \Phi^{cl}, \dots)$.

This is, furthermore, our \textit{prescription} on which are the states in the gauge field theory defined on $S^d \times (time)$ 

that correspond to a connected aAdS spacetime with some classical configuration of fields.

Below, we will give some arguments that support this claim from a basic set up for quantum gravity.

\vspace{.4cm}
\textbf{Holographic states from gravity}
\vspace{.4cm}

In order to better understand the mechanism of the holographic emergence, one should formulate the question by asking in which cases one has a well defined smooth aAdS geometry in the duality. 
If one starts from the strictly quantum gravity (QG) point of view, in the path integral formalism \cite{HH}, the states above are those that fulfill the following two basic conditions:
\begin{itemize}
\item[\textbf{(i)}] in the semiclassical limit, one recovers (smooth) classical fields and spacial geometries %as For the $\phi$-sector, it expresses as:
\be\label{expectation}
\langle\Psi^{\phi_{-}} |\widehat{\Phi} |\Psi^{\phi_{-}}\rangle = \Phi^{cl}(\phi_{-} , \phi^\star_{-}) + ``quantum \,\, correction''
\ee
for all the quantized fields of the gravitational theory  (matter fields + metrics)\footnote{Nevertheless, as explained above, we only study the $\phi$-sector for simplicty.}. The term ``quantum correction'' stands for any contribution that in AdS/CFT, typically goes as $\hbar/N^2 \to 0$ and $\Phi^{cl}$ denotes the (euclidean) classical solution, depending only on the boundary data $\phi_{-}$, and $\phi^\star_{-}$, which is defined from $\phi_-$ by reflection at the euclidean time $\tau =0$ (see \cite{us} and eq. \eqref{wavef-g-dual} in Appendix).
\item[\textbf{(ii)}] And furthermore (conversely), given \textit{any} classical bulk solution $\Phi^{cl}(\phi_{-} , \phi^\star_{-})$ one has the corresponding state \eqref{wavef-g}, and then \eqref{wavef-g-largeN} in the semi classical limit.\\
\end{itemize}

These are the more basic assumptions one might do in order to recover the notion of classical fields, on classical geometries, as univocally associated to a quantum state.

Since the connection between \eqref{estadoinicial-resultado} and \eqref{wavef-g} has already been clearly established in \cite{us}, based on the SvR approach \cite{SvRC,SvRL},
in the Appendix we will show that states defined as \eqref{wavef-g} satisfy these more basic criteria in a path integral formulation of (aAdS) QG.

The following, rather than a basic condition, is a property that typically present the states prescribed above. It will be directly  noticed from our anaysis in the next Section.
\begin{itemize}
\item[\textbf{(iii)}] These states are sufficient (form a basis) to describe certain, pertubatvely defined, (QG) space of states.  
\end{itemize}

\vspace{.4cm}
\textbf{The Large-N limit and quantum coherence}
\vspace{.4cm}

Taking the large-N limit in the above expression \eqref{wavef-g}, the Euclidean action for $\Phi$ becomes gaussian \cite{BDHM} and it also decouples from gravity, since the Newton constant $G_N \sim 1/ N^2$.

For free fields, another standard prescription \cite{BDHM,BDHM2,Kaplan} identifies the dual CFT operators with the boundary value of a canonically quantized field in AdS, $\hat{\Phi}(t,r,\Omega) $, as
\be
\label{kaplan}
\hat{ \mathcal{O}}(t,\Omega) \equiv\lim _{r\to\infty} \,r^{\Delta}\,\hat{\Phi}(t,r,\Omega)  =
\sum_k  a_k^\dagger F^*_k(t,\Omega) + a_k F_k(t,\Omega)\,,
\ee 
where of $F_k(t,\Omega) \equiv \lim _{r\to\infty} r^{\Delta}f_k(t,r,\Omega) $
defines a basis of functions on the conformal boundary and $f_k$ are the basis of normalizable solutions of the e.o.m. $( \Box - m^2 ) \Phi = 0 $ labeled by $k$, and $m^2=\Delta(\Delta-2)$.

Demanding consistency between both prescriptions, we
conclude that in the large-N limit, the state \eqref{estadoinicial-resultado} is \emph{coherent} and can be represented  as
 \begin{equation}\label{coherente-st}
 |\Psi^{\phi_{-}} \rangle  \propto e^{\sum_k \, \lambda_k \, a_k^\dagger} |0 \rangle\,,
\end{equation}
in the (bulk) Hilbert-Fock space ${\cal H}_{AdS}$. The same analysis is possible for linear perturbations
of the background metrics $g_{\mu \nu}$. Notice that these form a overcomplete basis, as an explicit realization of property \textbf{(iii)}.

This expression has been checked through explicit holographic computations of
correlation functions and inner products \cite{us}. These states are \emph{eigenstates} of the annihilation operators $a_k$ are their eigenvalues are precisely given by 

\be\label{lambdak} \lambda_k = -\int_{(-\infty,0] \times S^d}  \, d\tau d\Omega \, F^*_k(-i\tau,\Omega) \,\phi^-(\tau,\Omega)\,.\ee

On the other hand, the (excited) states $|E_k\rangle_1 \otimes |E_k\rangle_2$ of the basis in which the Black Hole entangled state is expanded in the Van Raamsdonk setup, correspond to eigenstates of the Hamiltonian operator 
 $H= H_\Phi + H_{gravity} + \dots + O(1/N)$, where
\be\label{hamcan}
: H_\Phi :=  \sum_{k} \epsilon_{k} \,a_{k}^{\dagger} \, a_{k}^{} \,\,,
\ee
and $\dots$ denote terms associated with other (matter) fields.

So there are states in the basis that express as $|E_k\rangle_1 \sim (a_{k}^{\dagger})^n |0\rangle_\phi \otimes |0\rangle_{gravity}\otimes |\dots\rangle$, where $|0\rangle_{gravity}$ denotes the exact globally AdS spacetime, or any other aAdS solution of vacuum Einstein equations\footnote{As argued above, this state can also be associated to some other background (euclidean) metric that minimizes the gravitational action, and then give wave functions \eqref{estadoinicial-resultado} in the large-N approximation.}. 
The key point in our argument is that this type of states (except the vacuum) are not eigenstates of $a_k$ since these operators do not commute with $H_\Phi$. Therefore, these states cannot be associated to classical configurations/solutions in the sense explained above.
In other words, there are states $|E_k\rangle_1 \otimes |E_k\rangle_2$ in the CFT$_1$ $\otimes$ CFT$_2$ Hilbert space which \emph{are not} dual to any classical spacetime with classical fields.

Let us remark that the states 
\eqref{coherente-st}, belonging to the bulk Fock space,
in fact fulfill the assumptions (i) and (ii), consistently with our expectation for states with a good holographic dual. This can be directly checked from eq. (\ref{coherente-st}), using (\ref{lambdak}), that:
\be\label{classical value}
\langle \Psi ^{\phi_{-}}|\,\widehat{\Phi} \,|\Psi^{\phi_{-}}\rangle = \Phi^{cl}(\phi_{-} , \phi^\star_{-})\,\,\,,
\ee
which agrees with \eqref{expectation}. This is satisfied for coherent states but fails for states $(a_{k}^{\dagger})^n \,|0\rangle_\phi $ with $n>0$. Moreover, for coherent states
 this property can be generalized to any functional ${\mathcal F} [\widehat{\Phi}]$ of the quantized fields and its derivatives (e.g. the Hamiltonian $H_\Phi$), which, with the proper normal ordering, coincides with the same functional valued on a classical solution:
\be
\langle \Psi^{\phi_{-}} | :  {\mathcal F} [\widehat{\Phi}]  : |\Psi^{\phi_{-}}\rangle = {\mathcal F} [\Phi^{cl}]  \,.
\ee

\vspace{.4cm}
\textbf{The eternal black hole as coherent state}
\vspace{.4cm}

Since the eternal AdS Black Hole is a geometric state of the doubled CFT theory, schematically denoted as CFT$^2$ ($\equiv$ CFT$_1\,\otimes$ CFT$_2$ ) \cite{collapse}, this indeed fits into the proper extension of the notion of coherent state.

Thermal Bogoliubov's transformations $G_\beta$ are formally unitary and canonical in the sense of preserving the canonical commutation relations and the norm of the states in finite volume systems,
so then, the thermal TFD state (dual to a AdS black hole) \eqref{BHstate} is an eigenstate (with eigenvalue $=0$) of the thermal annihilation operators
\be
a_k^{ \; (1)}(\beta) \equiv  a^{ (1)}_k - e^{-\beta \epsilon_k /2} a^{\dagger \; (2)}_k = G_\beta \, a^{ \; (1)}_k \, G^\dagger_\beta
\ee
\be
a_k^{ \; (2)}(\beta) \equiv  a^{ (2)}_k - e^{-\beta \epsilon_k /2} a^{\dagger \; (1)}_k = G_\beta \, a^{ \; (2)}_k \, G^\dagger_\beta\,
\ee
then, the equations $a_k^{ \; (i)}(\beta) |0(\beta)\rangle = 0 $, $i=1,2$ are solved by the state \cite{collapse}:
\be \left|0(\beta)\right\rangle_{(\Phi)} =
G_\beta\,  \left|0\right\rangle = Z_\Phi^{-1/2}
\prod_{k}\left[
e^{(e^{-\beta \epsilon_k /2})\,\,a_{k}^{\dagger\:(1)}
a_{k}^{\dagger\:(2)}} \right]
\left|0\right\rangle \label{BHstate-bulk}. \ee
This is the state \eqref{BHstate} represented in the scalar field sector of doubled AdS Hilbert-Fock space, in the same sense that the states \eqref{coherente-st} in a single CFT theory.
This state is also invariant under the action of the extended Hamiltonian $H^{(1)}_\Phi - H^{(2)}_\Phi$ \cite{tu,ume1,ume2}.
%\newpage

\vspace{.3cm}
\textbf{Holographic excitations with two conformal boundaries}
\vspace{.3cm}

\begin{figure}[t]\centering
\begin{subfigure}{0.49\textwidth}\centering
\includegraphics[width=.9\linewidth] {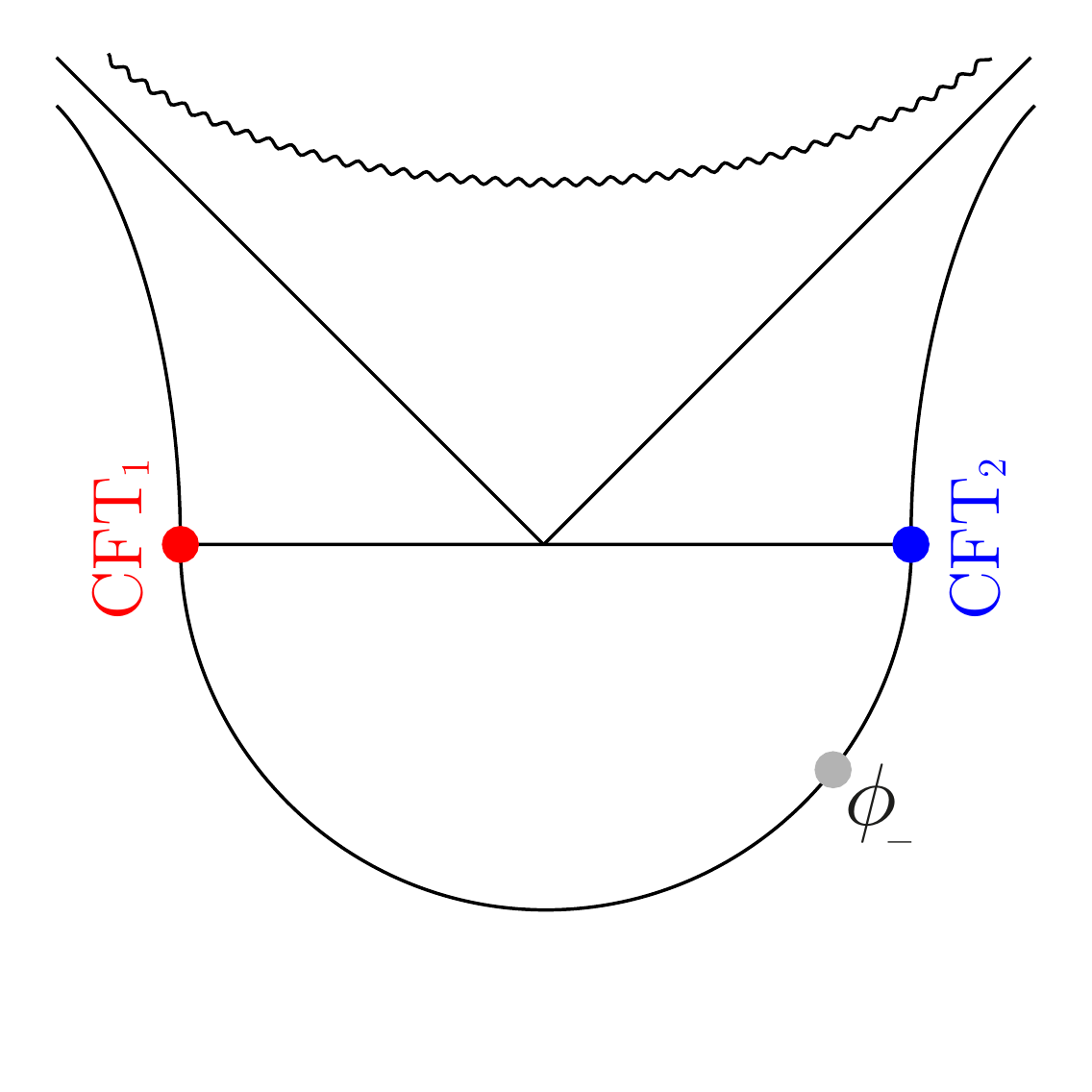}
\caption{}
\end{subfigure}
\begin{subfigure}{0.49\textwidth}\centering
\includegraphics[width=.9\linewidth] {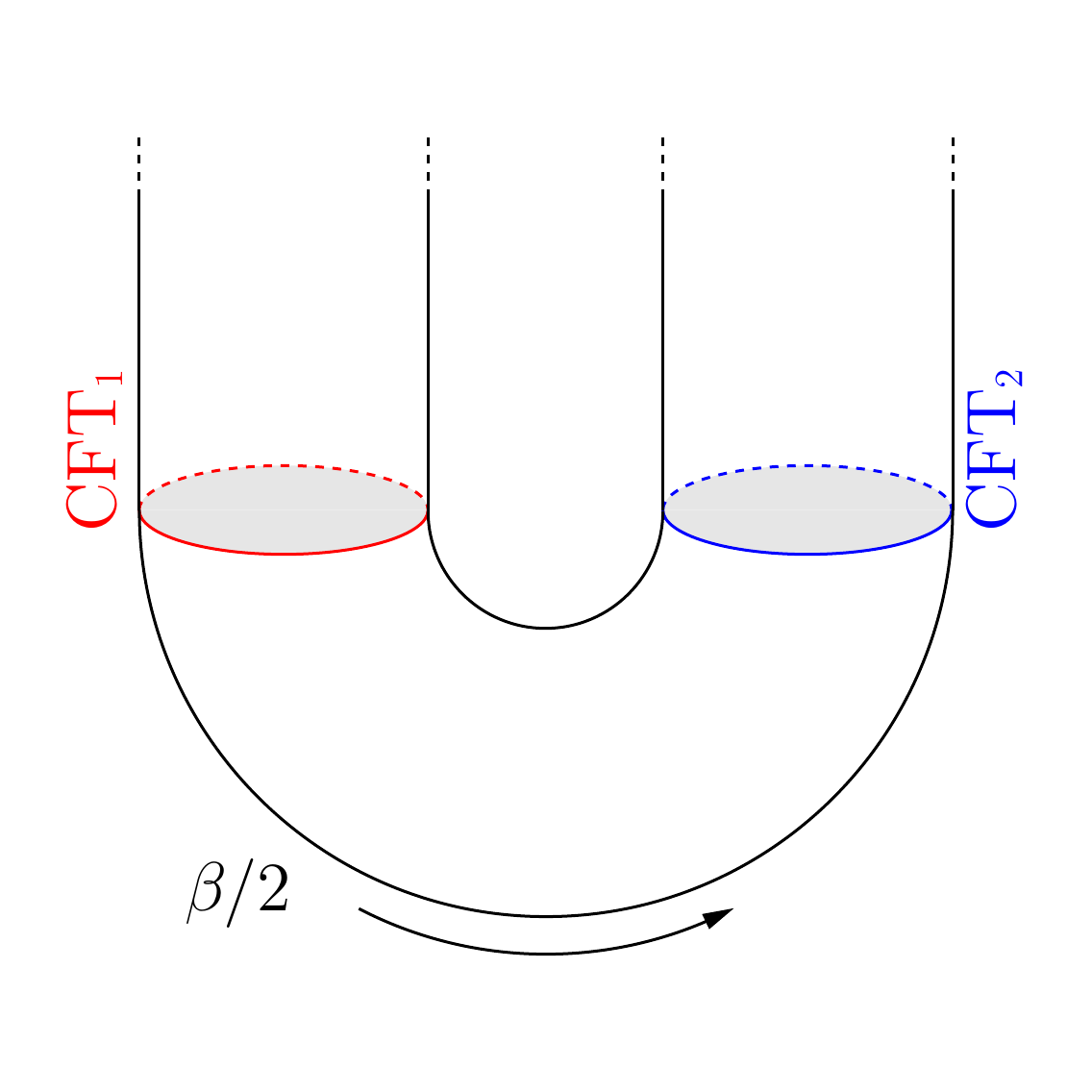}
\caption{}
\end{subfigure}
\caption{\small{(a) It illustrates the geometry dual to the state \eqref{estadoinicial-BH}, where the semi-circle represents $\partial_r {\cal M_{BH}}_- = [0, \beta/2]\times S^d $ with a source $\phi_-\neq 0$. The euclidean section continues into a Lorentzian evolution with two causally disconnected asymptotic boundaries. (b) At low temperatures ($\beta\to \infty$ limit) the correlations between both CFTs vanish, and the dual geometry can be described as two totally disconnected copies of AdS.}}
\label{BH+HP}
\end{figure}

In this case our prescription (\ref{estadoinicial-resultado}) for the holographic states expresses as
\be\label{estadoinicial-BH}
 | \Psi^{\phi_-}(\beta) \rangle = \, e^{-\int_{\partial_r {\cal M_{BH}}_-}  \, \mathcal{O} _\beta\, \phi_-} \,  |0(\beta) \rangle\,.
\ee
where ${\cal M_{BH}}_-$ denotes the (lower) half of the Euclidean black hole (Fig \ref{BH+HP}(a)), and $\mathcal{O}_\beta \equiv\mathcal{O}^{(1)}_\beta(x,\tau)$ represents a local operator on $\partial_r {\cal M_{BH}}_- = [0, \beta/2]\times S^d $, which is defined from the CFT$_1$ operators by analitical extension to the Euclidean (past) time. Similiarly, another set of independent (completely commuting with $\mathcal{O}^{(1)}_\beta$) operators, $\mathcal{O}^{(2)}_\beta$ can be defined from CFT$_2$.

By following the procedure used in \cite{us}, quantizing the \emph{probing} free field in this bulk and using the standard BDHM prescription \cite{BDHM}, one can see that the holographic states correspond to nothing but thermal coherent states\footnote{This confirms the guess on which should be the states of CFT$^2$ with geometric dual, pointed out in \cite{collapse}}:
\be\label{BH-excitations}
|\lambda(\beta)\rangle \equiv D(\lambda )|0(\beta)\rangle = e^{\lambda_k \,a_k^\dagger (\beta) -\lambda_k ^* \, a_k(\beta)} |0(\beta)\rangle
\ee
where the index $(i)=1,2$ is implicit and $D(\lambda)$ is the corresponding displacement operator which generates the coherent state  $D(\lambda)|0\rangle =|\lambda \rangle$. 
The formula for $\lambda_k$ is similar to \eqref{lambdak} but the integration is on the euclidean interval $[0, \beta/2]\times S^d$ and the functions $F^{\, *}_k(\tau, \Omega)$ correspond to the normalizable modes of the AdS-Schwarschild solution through the BDHM prescription \cite{BDHM,BDHM2}. 
An extended study on this excitations in AdS black holes and a detailed proof of the results mentioned here are being prepared and will be presented in a forthcoming work \cite{us3}.

This extends our prescription (\ref{estadoinicial-resultado}) of states of the theory CFT$^2$  with good classically connected spacetime interpolating between two (disconnected) asymptotic boundaries $S^d \times \Re$.

Noticeably, these states/geometries have a similar description that the Black Hole, and decompose as in Fig. 1 although the coefficients appearing in Fig 1(b) change accordingly.

\vspace{.4cm}
\textbf{A brief comment on the Hawking-Page quantum transition}
\vspace{.4cm}

Let us finally point out an implication of our arguments on the Hawking-Page phase transition \cite{haw-page1,haw-page2}.  It can be be described in a quantum mechanical way as a critical behavior of the quantum amplitude and standard quantum collapse \cite{collapse}, in line with the Van Raamsdonk interpretation. Since the AdS-BH spacetime is described by the state (\ref{BHstate}) in CFT$^2$ (figures \ref{vanramfig}a and \ref{BH+HP}a) ; which at low temperatures (compared with the AdS scale $\beta\sim R_{AdS}$) reads as
\be \label{BHstate-low}
|0(\beta)\rangle =  \,\frac{e^{-\frac{\beta}{2} E_0}}{Z^{1/2}} | 0 \rangle_1 \otimes | 0 \rangle_2 + \sum_{n\neq 0} \delta_n (\beta) |E_{n}\rangle_1 \otimes |E_{n}\rangle_2\,\,\,, \ee
where $|E_{n}\rangle$ are states orthogonal to $|0\rangle_1 \otimes |0\rangle_2$, and $|\delta_k(\beta)|^2 \ll |Z^{-1}\,e^{-\beta E_0}|$. as $\beta > E_0^{-1} \sim R$ (here $R$ is the radius of curvature of the AdS space).
So at low temperatures, the probability of collapsing it, and to ``observe'' two disconnected global AdS spaces (see fig. \ref{BH+HP}b) is very high, compared with other states. 

However this description must be complemented with our previous arguments since it should be explained why, for the high temperature regime where the amplitudes are comparable, the system cannot be observed in some of the excited states $|E_{n}\rangle_1 \otimes |E_{n}\rangle_2$ upon measurements, giving place to other geometric transitions. The reason might be that such (micro)states would not be compatible with direct observations or measurements of local fields on a spacetime.
The only holographic states \eqref{estadoinicial-resultado} with probing fields that vanish assymptotically, precisely are the black hole state and the ground state\footnote{The state $|0\rangle_1 \otimes |0\rangle_2$ here only expresses the gravity sector. For simplicity, we have omitted the other (thermal) excitations: of the matter fields that include the $\Phi$-normalizable modes.} $|0\rangle_1 \otimes |0\rangle_2$. 
Some aspects of the present description of the transition \cite{collapse}
 are shared by the approach of \cite{Jafferis}.

\vspace{.4cm}
\textbf{Conclusions}
\vspace{.4cm}

Here, we have studied \emph{how} a classical geometry holographically corresponds to a CFT state, and prescribed which these states should be. In fact, using basics tools of holography as the SvR prescriptions %\footnote{Which is the natural extension to real time of GKPW} 
and the HH construction for QG, we have
explicitly shown that this is a one to one map between such CFT states and classical (euclidean) geometries.

As evidence we have explicitly shown that the $d+1$ dimensional spacetime geometries, as probed by non-back reacting free fields, corresponds to coherent states dual to CFT states as \eqref{estadoinicial-resultado} defined on the conformal boundary. In summary, the proposed holographic CFT states are univocally characterized by the boundary conditions on a half ($\tau < 0$) of the radial asymptotic boundary of the euclidean spacetime: $\left| \Psi^{\phi_-} \right\rangle$.
As a first consequence of this, we observed that the basis microstates of the linear superposition \eqref{BHstate-bulk} should not be directly interpreted as classical aAdS spacetimes in the gravity side, but in a non standard way, as quantum superposition 
of well defined geometries

Another different family of asymptotically AdS spacetimes are obtained by entangling the CFT with another identical (non-interacting) copy \cite{vanram, collapse}, whose microstates are also coherent states (see for instance, eqs. \eqref{BHstate-bulk}, and \eqref{BH-excitations}).

Interestingly, we would like to stress that holographic states with many component boundaries as the BH or its excitations, have at least two different geometric dual descriptions. It is fascinating that they describe two very different spacetime topologies.
One of these representations is continuous and classically connected spacetime as in the states \eqref{estadoinicial-BH} whose dual are slightly deformations of an eternal BH geometry, and the other one, looks like schematically as the superposition of Fig 1(b), although as we observed in the present work, the basis microstates should be products of coherent states. 

An intriguing question that arises here is if (and why) one of these dual geometric representations (in terms of connected or disconnected geometries) should be ``preferred" by the system. Furthermore, since the coherent basis of holographic states is overcomplete, how should be interpreted the quantum collapse into some specific element/geometry of this basis.

\vspace{.4cm}
{\small \textbf{Acknowledgements} The authors are grateful to CONICET for financial support. Thanks are due to Guillermo Silva for
useful discussions, specially in the context of \cite{us3}}.

\vspace{.4cm}

{\small\textbf{Note added:} The first version of this work was written as an essay for the Gravity Research Foundation 2016 Awards for Essays on Gravitation. In the meantime, since the first version appeared in Ar$\chi$iv until the present submission, appeared \cite{vanram2} with certain overlap on our claim around the formula \eqref{estadoinicial-resultado} for holographic states. In ref. \cite{vanram2} the authors checked, through a perturbative construction, that these states have a good dual aAdS spacetime}.

\vspace{.4cm}
\textbf{Appendix}
\vspace{.4cm}

\emph{Proof of \textbf{(i)}}
Start by defining a family of states of QG (in the path integral formulation \cite{HH})
by the wave functionals \eqref{wavef-g} projected on a complete bulk configuration basis $\phi_{\Sigma}$ on $\Sigma^{}$, let us refer to this as ${\cal H _P}$. 
Its corresponding dual $\Psi^{\phi^\star_{-}}[\phi_{\Sigma}] \in {\cal H}_{\cal P}^*$ was defined in ref \cite{us} as the path integral
\begin{equation}
\label{wavef-g-dual}
\left(\Psi^{\phi_{-}}[\phi_{\Sigma}]\right)^\star \equiv  \Psi^{\phi^\star_{-}}[\phi_{\Sigma}] = \int [\mathcal{D}\Phi]_{(\phi_{\Sigma},\phi^\star_{-})} \,  e^{-S^{}_+[\Phi]}\,\, ,
\end{equation}
where $\phi^\star_{-} $ is defined from $\phi_{-}$ on the hemisphere $\partial^+ M $, by (euclidean) time reflection in the equator $\tau=0$, that glues to real time at $t_\Sigma$ (see fig. \ref{SvRb} or \cite{us} for more details); $S^{}_+[\Phi]$ denotes the action valued on euclidean manifolds anchored by the boundaries $\partial^+ M $ and $\Sigma$. Recall that these states are unnormalized.
Then, for instance, the expectation value of the field operator is
\begin{align*}
\langle \Psi^{\phi_{-}} |\widehat{\Phi}(x,t_\Sigma) |\Psi^{\phi_{-}}\rangle &= \sum_{\phi_\Sigma } \langle \Psi^{\phi_{-}} |\widehat{\Phi}(x,t_\Sigma) |\phi_\Sigma (x)\rangle\langle\phi_\Sigma (x)|\Psi^{\phi_{-}}\rangle \\
&= \sum_{\phi_\Sigma } \,\,\phi_\Sigma \,\,\Psi^{\phi^\star_{-}}[\phi_{\Sigma}] \Psi^{\phi_{-}}[\phi_{\Sigma}] \\
&= \int [\mathcal{D}\Phi]_{\phi^\star_{-}, \phi_{-}} \,\phi_\Sigma(x)\,  e^{-S^{}[\Phi]}\,\, ,\nonumber
\end{align*}
where we have used that the configuration basis $|\phi_\Sigma (x)\rangle $, $\phi_\Sigma(x)\equiv \phi(x, t_\Sigma)$, is complete and diagonalizes the operator $\widehat{\Phi}(x,t_\Sigma)$. Therefore, in the saddle point approximation, the r.h.s reads
\be
 \int [\mathcal{D}\Phi]_{\phi^\star_{-}, \phi_{-}} \, \phi(x, t_\Sigma)\,  e^{-S^{}[\Phi]} \approx \Phi^{cl}(x, t_\Sigma) e^{-S^{0}[\phi^\star_{-}, \phi_{-}]} + O(\hbar/N^2)
\ee
%(Notice $\phi_\Sigma(x)\equiv \phi(x, t_\Sigma)$) 
where $ \Phi^{cl}(x, t_\Sigma)= \Phi^{cl}(x, \tau=0)$ is the (euclidean) classical solution for the boundary data $\phi^\star_{-}, \phi_{-}$. This realizes the condition \eqref{expectation}.

\vspace{.8 cm}
\emph{Proof of \textbf{(ii)}}
The computation above manifestly shows that given the state \eqref{wavef-g} one univocally recovers the classical configuration of the bulk fields on $\Sigma$ (including the metric). Reciprocally, given a classical solution $\Phi^{cl}(x, \tau)$ one can to define the boundary data:
\be
\phi_\Sigma(r,\textbf{x}) :=\Phi^{cl}(r,\textbf{x}, \tau=0)\,\,;\qquad
\phi^- (\textbf{x} , \tau) := \,\,\lim_{r\to \infty }\,\,\Theta(-\tau)\Phi^{cl}(r,\textbf{x}, \tau)
\ee %\phi^0(x, \tau < t_\Sigma)
and then univocally one can define the state 
\eqref{wavef-g}.
 Therefore, the map $\{ \Psi^{\phi^-} \} \leftrightarrow \{\Phi^{cl} [\phi^\star_{-}, \phi_{-}]\}$ is bijective.

\end{document}